\documentstyle[12pt]{article}
\def\be{\begin{equation}}
\def\ee{\end{equation}}
\def\bea{\begin{eqnarray}}
\def\eea{\end{eqnarray}}
\def\a{\alpha}
\def\b{\beta}
\def\e{\varepsilon}
\def\g{\gamma}

\begin{document}
\renewcommand{\theequation}{\thesection.\arabic{equation}}
\title{Isotropic universe with almost scale-invariant fourth-order gravity}
\author{Hans-J\"urgen  Schmidt and Douglas Singleton}
\date{May 13,  2013}
\maketitle
\centerline{Institut f\"ur Mathematik, Universit\"at Potsdam, 
Germany\footnote{Permanent address of DS: Physics Department, California 
State University, Fresno CA 93740 USA,  dougs@csufresno.edu,
current address of DS: Dept. Physics, Institut Teknologi 
Bandung, Indonesia}} 
\centerline{Am Neuen Palais 10, D-14469 Potsdam,   hjschmi@rz.uni-potsdam.de}
\begin{abstract}
We study a class of isotropic cosmologies in fourth-order gravity with 
Lagrangians of the form $L= f(R) + k (G)$ where $R$ and $G$ are the Ricci and 
Gauss-Bonnet scalars  respectively. A general discussion is given on the conditions under 
which this gravitational Lagrangian is scale-invariant or {\it almost} scale-invariant. 
We then apply this general background to the specific case 
$L  = \alpha R^{2} +  \beta \, G \ln G$ with constants $\alpha, \beta$. 
We find closed form cosmological solutions for this case.  One interesting feature 
of this choice  of  $f(R)$ and $k(G)$ is that for very small negative  value of the 
parameter  $ \beta$ the Lagrangian   $L= R^2/3 + \beta G \ln G$ leads to the replacement 
of  the exact de Sitter solution coming from $L=R^2$ (which is a local attractor)  to 
an exact, power-law inflation solution $a(t) = t^p = t^{-3/\beta}$  which is also a local 
attractor.  This shows how one can modify the dynamics from de Sitter to power-law 
inflation by  the addition of  an $G \ln G$-term.
\end{abstract}


\noindent  \hspace*{2cm} scale-invariant  gravity,  fourth-order gravity

\noindent AMS-Classification: 83D05, 83F05, 83C15, 53Z05, 85A40

\section{Introduction}
\setcounter{equation}{0} 
From the huge class of theories of gravitation which can be considered for describing
and explaining the early evolution of the Universe, it is the subclass of scale-invariant 
ones which plays a prominent role. The reason for this prominence is that almost 
all physical theories and their resulting cosmologies have some limiting regime which
 is free from any scales. \par
In the present paper we investigate a  class of  cosmologies which include
 not only scale-invariant theories but also {\it almost} scale-invariant theories. 
 In section 2 we will present concrete  definitions and results related to the 
 several variants of  ``(almost) scale-invariant theories".  \par
The Lagrangians which we consider are of the form
\be 
 L = f(R) + k(G),
\ee
 where $R$ is the curvature scalar and  
$$
G = R_{ijkl}R^{ijkl} - 4 R_{ij}R^{ij} + R^2 
$$
is the Gauss-Bonnet scalar. By placing restrictions on the form of the functions
 $f(R)$ and $k(G)$ we will obtain theories which are scale-invariant and almost 
scale-invariant. We will focus on cosmological solutions to these scale-invariant and  
almost scale-invariant Lagrangians  -- in particular spatially flat Friedmann 
space-times. We obtain general features for these almost scale-invariant cosmologies,
 and for certain cases we are able to completely integrate the resulting Friedmann-like
 equations to confirm older results and to  obtain  new exact, closed form solutions 
 in terms of the Friedmann metric scale factor, $a(t)$. \par
 Before moving to the detailed calculations  we give a brief review of work in this area 
which has some connection with the present  paper. 
 Cosmological models where the action depends on  the Gauss-Bonnet 
scalar  $G $ are discussed in  \cite{s8},  \cite{s12},   \cite{s13},  and  \cite{s9}. 
 In  \cite{s14}, exact solutions for $k(G) = G^\beta$ are given which have an 
ideal fluid  source, a power law scale factor, $a(t) = t^p$, with $p$ depending 
on $\beta$  and  the equation of state of the fluid; $a(t)$ is the  related cosmic 
scale factor. Further papers on this  topic are  \cite{s14a}, 
  \cite{s15}, \cite{s21}, \cite{s52}, \cite{s53},  \cite{s19},  \cite{s60}, 
 \cite{s50}, and  \cite{s62}. In \cite{s994}, the  $\Lambda$CDM epoch 
reconstruction from $F(R,G)$ and modified Gauss-Bonnet gravities is presented.
In this work models with  Lagrangians $R + k(G)$, or more general $f(R) + k(G)$, 
and also $R + \xi(\phi) G + \phi_{,i} \phi^{,i}$ are discussed especially for the 
spatially flat Friedmann models.   \par 
The paper  \cite{s995} investigates $\Lambda$CDM cosmological  models, using 
Lagrangians of the form $L=k(G)$, and $L=R + k(G)$.  For the pure $k(G)$ gravity, 
and a spatially flat Friedmann model with scale factor $a(t)$ where $t$ is synchronized 
time, the following results are obtained: A de Sitter space-time with Hubble parameter 
$h=\frac{\dot a}{a} >0$  has $G = 24 h^4$. This de Sitter solution is a vacuum 
solution if the condition  $G dk/dG = k(G)$ is fulfilled. The exact power-law 
solution of the form $a(t) =t^p$  exists if  the following condition is fulfilled:
\be 
0 = G \frac{dk}{dG} - k(G) + \frac{4G^2}{p - 1} \cdot \frac{d^2k}{dG^2}, 
\ee
i.e. if $k(G) = G^{(1-p)/4}$, or more completely the Euler-type 
eq. (1.2)  has solutions $k(G) = c_1 G + c_2 G^{(1-p)/4}$ with constants 
$c_1$ and $c_2$ -- compare with eqs. (59) and (60) of   \cite{s995}. The term
$c_1 G$ is a divergence, and so does not contribute to the field equation,
so, seemingly, only power-law Lagrangians  $k(G) = G^{(1-p)/4}$ produce 
the  exact solution $a(t) = t^p$. However, this is not the complete truth: 
If one takes the example $p = -3$, then the solutions of eq. (1.2)  become
$k(G) = c_1 G + c_2 G \ln G$, a case not mentioned in    \cite{s995}. 
So, besides powers of $G$, also $G \ln G$ leads to exact solutions $a(t) = t^p$. 
 Further recent papers on $k(G)$ gravity are  \cite{s22},  \cite{s49}, \cite{s20}, 
 and \cite{s74}. In \cite{s1}, the Lagrangian
\be 
L = G \ln G
\ee
is discussed, and cosmological closed-form solutions are given, 
including the just mentioned exact solution $a(t) = t^p$ with $p=-3$. 
\par
In  \cite{s997}, the anomalous velocity curve of spiral
galaxies is modelled by Lagrangians of type $L(R, G, \Box G)$, where
$\Box$ denotes the D'Alembertian, especially in the form 
\be 
L= \tilde G \ln \tilde G, \qquad {\rm where} \qquad \tilde G = \frac{G}{
(\Box + \a R)G}. 
\ee
In a first approximation, one can assume $\tilde G \approx G$, so this
Lagrangian has similarity with that one from eq. (1.3).  \par
In \cite{s998}, the  stability of power-law solutions in cosmology is discussed
for $L=G$, which gives non-trivial results for space-time dimension exceeding 4 only.
 In    \cite{s69},  solutions for  $L = R + \sqrt{G} $ with  a Friedmann scale factor 
 of power-law form, i.e. $a(t) = t^p$  are given. In   \cite{s76}, the 
 Lagrangian  $L = R^n + \beta G^{n/2}$ is investigated. Further Lovelock models 
along the  line  of   \cite{s76}  are given in    \cite{s82},  whereas in  \cite{s83}  the
 case  $k(G) = G^n + \b G \ln G$   is discussed. \par
In \cite{s996},  the stability of the cosmological solutions  with matter 
 in $f(R,G)$ gravity  is discussed,  with special emphasis on the stability of the 
de Sitter solution, and with Lagrangians of the type $R + R^nG^m$.   \par
Analogous models for $f(R)$-gravity can be found in  \cite{s70x}, where the case
 $L = R^{3/2}$ is related to Mach's principle. In  \cite{s70y} an exact solution
 for $L=R^2$  is given. Further models are discussed in  \cite{s70}, \cite{s70a}, 
\cite{s70b}, \cite{s70c}, \cite{s5}, and \cite{s18}. In \cite{s992}, the stability of 
models within theories of type $L=R + R^m + R^n$ with $n<0<m$ is discussed, 
and exact power-law solutions are obtained.  Further papers on this topic are
 \cite{s57},  \cite{s58},  \cite{s56}, \cite{s55},  and \cite{s2}. In  \cite{s72}, 
  the case  $f(R) = R^{3/2}$ is studied.  In \cite{s993}, the following strict result
 is shown: Exact  power-law cosmic expansion in $f(R)$ gravity models with perfect 
fluid as source is possible for $f(R)=R^n$ only.  Newer models of this kind can be 
found in  \cite{s46},  \cite{s25}, \cite{s26}, \cite{s27},  \cite{s28}, \cite{s29},
 \cite{s30},  \cite{s33},  \cite{s36}, \cite{s37}, \cite{s38}, \cite{s39},  \cite{s41}, 
\cite{s42},   \cite{s43},  \cite{s6}, \cite{s7}, \cite{s45},  \cite{s47}, \cite{s44}, 
\cite{s48},  \cite{s65},  \cite{s66},  \cite{s32},  \cite{s35}, \cite{s71}, \cite{s40}, 
\cite{s79},  \cite{s63},  \cite{s11},  \cite{s24},  \cite{s77}, \cite{s64}, \cite{s67}, 
\cite{s75},  \cite{s81}, and  \cite{s23}.  \par
 The conformal Weyl theory, especially the value of the perihelion advance
 in this theory, has been discussed in  \cite{s61},  \cite{s54},    \cite{s3}, \cite{s16},   
 \cite{s31},    \cite{s4},    \cite{s78},    \cite{s80},    \cite{s34},  and  \cite{s68}. 
 For theories in lower-dimensional space-times see e.g. \cite{s999}, 
 \cite{s51},   and  \cite{s10}. \par
Our motivation for considering Lagrangians of the form given in eq. (1.1) is as 
follows: We study the cosmological aspects of a specific version of   $F(R,G)$
 gravity which is scale-invariant in the sense that in the absence of matter no 
 fundamental length exists within that theory. One can contrast this with 
$R \pm  l^2 R^2$  theories which have the fundamental length $l$. In connection 
with this we discuss and clarify that there are slightly different notions of 
scale-invariance, and we carefully distinguish between them. We do not insist on 
second-order field equations,  \footnote{Of course,
models with $G$ entering  linearly  the Lagrangian became popular, because  
in dimensions larger than 4,  they lead to  second-order field equations, but this 
is not our goal here.}  so also non-linear 
dependences of the Lagrangian on $G$ are included with the result, that the field 
equations are of fourth-order in general. Similarly, we do not motivate our 
research by  string  theory,\footnote{Of  course, others   \cite{s8}
 mention just the connection with string theory as motivation.}  but rather we 
want to present  possible models  for the observed evolution of the universe
which includes both inflationary phase at early times and the present acceleration
(normally attributed to some fluid/field generically termed ``dark energy")
 without the  need to introduce additional matter fields.
Thus our motivation is as follows: First the leading principle is
that in the first approximation, the Einstein-Hilbert Lagrangian $R$ is the right
 one for weak fields. Second, a non-linear addition to the Einstein-Hilbert Lagrangian 
 depending on $R$, especially of the form $R^2$  or  $R^2 \ln R$, gives the desired 
 early time inflationary behaviour, see e.g. the early papers    \cite{s70} on this topic.
 Third, the further addition of a term non-linear in $G$ to the Lagrangian
 was proposed in   \cite{s992},  \cite{s13}  and others as a possible alternative 
 for dark energy, which is the generic term for the substance postulated to drive 
 the current accelerated expansion of the Universe. To make the task tractable of 
finding which of these various modified gravity theories can give the observed
 late time acceleration the authors of
 \cite{s14a} developed ``the reconstruction program for the number 
of modified gravities including  scalar-tensor theory, $f(R)$, $F(G)$ and 
string-inspired, scalar-Gauss-Bonnet gravity. The known (classical) universe 
expansion history is used for the explicit and successful reconstruction of some 
versions (of special form or with specific potentials) from all above modified gravities." 
\par
The paper is organized as follows: As already said, in section 2 we give a general 
discussion of scale-invariance and almost scale-invariance. This general discussion 
motivates our  special choice 
\be 
L  = \alpha R^{2} +  \beta G \ln G.
\ee
for the Lagrangian. Section 3 gives a brief, self-contained review 
of relevant formulas concerning the Gauss-Bonnet scalar and Gauss-Bonnet gravity. 
This is done since $k(G)$ models are much less  known that $f(R)$ models.
Section 4 gives our main new results which follow from  the almost scale-invariant 
Lagrangian of the form eq. (1.5). Section 5 summarizes and gives conclusions 
about the results presented in this  paper. 

\section{Notions of   (almost)  scale-invariance}
\setcounter{equation}{0} 

 We first give the exact definitions of what we mean by an (almost)  scale-invariant 
gravitational  action or gravitational Lagrangian. A theory of gravitation with a 
geometric  Lagrangian  $L=L(g_{ij}, \partial_k)$ is defined by a scalar $L$ which 
 depends  on the metric and its partial derivatives up to arbitrary order. The 
 signature of the metric is $(-+ \dots +)$ and $g = \det g_{ij}$. Within this section, 
we assume the dimension of space-time to be $D \ge 2$. Then the gravitational action
$I$ is defined by  
\be 
I = \int L \sqrt{-g} d^D x. 
\ee
A scale-transformation, also called a homothetic transformation, is a conformal 
transformation with a constant conformal factor. In another context, 
 scale-transformations can also be interpreted as  transformations that change the 
applied length unit. The transformed metric, ${\tilde g}_{ij}$, is defined as
\be 
\tilde g_{ij} = e^{2c} g_{ij}
\ee
where $c$ is an arbitrary constant.  For the inverted metric one gets
$$
\tilde g^{ij} = e^{-2c} g^{ij}.
$$
The Christoffel affinity $\Gamma^i _{jk}$, the Ricci tensor $R_{ij}$ and the 
Riemann tensor $R^i_{jkl}$ do not change under the scale-transformation given in 
eq. (2.2), and also all covariant derivatives of the Ricci and the Riemann tensor are
 homothetically invariant. However $R$, $G$ and $g$ are changed under the 
transformation of eq. (2.2.) as follows:
\be 
\tilde R = e^{-2c} R, \qquad \tilde G = e^{-4c} R, \qquad \tilde g = e^{2Dc} g.
\ee
{\it  Definition:} The action (2.1) is called   scale-invariant, if $\tilde I = I$ 
 according to eq.  (2.2).  It  is called almost  scale-invariant, if  the difference
$\tilde I - I$  is a topological invariant.  \par
The Lagrangian $L$ is called scale-invariant if  there exists a 
constant $m$, such that 
\begin{equation} 
\tilde L \equiv L(\tilde g_{ij} ) = e^{mc} L (g_{ij} ) .
\end{equation}  
Finally, the Lagrangian $L$  is called almost  scale-invariant, if  the difference
$\tilde L  - e^{mc} L$  is  a  divergence.  \par
Of course, the sum of a scale-invariant action  and of an arbitrary topological  invariant 
is always an almost scale-invariant action. Likewise, the sum of  a  scale-invariant 
Lagrangian and a divergence is  always an almost scale-invariant Lagrangian. At first glance 
one might be tempted to conclude that the converse should be true, that an almost 
scale-invariant action can always be written as the sum of a scale-invariant action plus a 
topological invariant and that an almost scale-invariant  Lagrangian can always be written 
as the sum of a scale-invariant Lagrangian plus  a divergence.  However, as we will show
 below, there exist non-trivial examples  of almost scale-invariant actions which 
cannot be represented in the form of such  a sum. \par
The following relations between these four notions of scale-invariance
exist: If $L$ is scale-invariant, then with eq. (2.3) we get
$$
\tilde I \equiv \int \tilde L \sqrt{- \tilde g} d^D x
= e^{(m+D)c}\int L \sqrt{-g} d^D x = e^{(m+D)c} I, 
$$
so for $m=-D$, a scale-invariant Lagrangian gives rise to a scale-invariant action. \par
Likewise for  $m=-D$, an almost scale-invariant Lagrangian gives rise to 
an almost scale-invariant action, because the space-time
integral of a divergence represents a topological invariant. \par
 Let us now take the example $L= f(R)$ with an arbitrary but sufficiently smooth 
function $f$ and ask, under which circumstances, this leads to    scale-invariance. 
We have to distinguish two cases: $D=2$ and $D>2$. For $D=2$, the scalar
 $R$  represents a divergence, whereas for $D>2$, no function of $R$ has such a 
property. \par  
Let us start with the more tractable case $D>2$. As no function 
of $R$ gives a divergence, 
the notions of scale-invariance
 and almost scale-invariance coincide. For $L=f(R)$ to
 be a scale-invariant Lagrangian
 there must exist an $m$ such 
that the following relationship holds
\be 
f( \tilde R) \equiv f( e^{-2c} R ) = e^{mc} f(R)
\ee
 using  eqs. (2.2), (2.3), and (2.4). With $f'$ denoting the derivative of $f$
with respect to its argument we get from eq. (2.5) by applying $d/dc$
$$
-2 f'(e^{-2c} R) \cdot e^{-2c} R = m e^{mc} f(R).
$$
Putting $c=0$ into this equation we get a differential equation for $f(R)$: 
\be 
-2 R f'( R)  = m  f(R)
\ee
which is solved by 
\be 
f(R) = c_1 \cdot R^{-m/2}
\ee
with integration constant $c_1$. As expected, just the powers of  $R$  lead to 
scale-invariant Lagrangians. The corresponding action $I$ turns out to be 
scale-invariant  for $m = -D$ only, i.e. $L= R^{D/2}$ leads to a scale-invariant action, 
for  $D=4$ this is the celebrated $L=R^2$. \par
Let us now turn to the less trivial case $D=2$, where $R$ represents a divergence.
 We look for the set of all almost scale-invariant Lagrangians.  For a Lagrangian of the 
 form $L=f(R)$ to be almost scale-invariant  requires that there exists an $m$ such that
\be 
f( \tilde R) \equiv f( e^{-2c} R ) = e^{mc} f(R) + v(c) \cdot R
\ee
where $v$ depends on $c$ only to ensure that $v \cdot R$ is a divergence for
every $c$. Applying $d/dc$ we now get 
$$
-2 f'(e^{-2c} R) \cdot e^{-2c} R = m e^{mc} f(R) + v'(c) \cdot R.
$$
Inserting $c=0$ and abbreviating $v'(0)$ by $c_2$ we get in place of eq. (2.6) now 
\be 
-2 R f'( R)  = m  f(R) + c_2 \cdot R.
\ee
We divide by $R$,   apply $d/dR$ and get 
\be 
-2 \frac{d^2 f}{dR^2} = \frac{d}{dR} \left(  \frac{m f}{R}  \right)
\ee
which is solved by 
\be 
f(R) = c_3  R + c_4  R^{-m/2}
\ee
with integration constants $c_3$ and $c_4$. This is just what one expected from the 
beginning: The divergence  $c_3  R$ added to the power-law term $ c_4  R^{-m/2}$, 
 i.e. the added divergence term $v(c) \cdot R$ in eq. (2.8)  leads to the extra divergence 
term $c_3 R$ in eq. (2.11). However, eq. (2.10) possesses a further solution besides 
eq. (2.11): For $m=-2$  eq. (2.10) is solved by
\be 
f(R) = c_3  R + c_4  R \ln R. 
\ee
The result of eq. (2.12) was already noted in  \cite{s999}: Besides what one 
would have expected, the action $ I= \int R \ln R \sqrt{-g} d^2x$ turns out to
 be  almost scale-invariant. \par
Now we perform the analogous analysis for the Lagrangian $L = k(G)$.  For dimension 
$D \le 3$, $G$ vanishes, so this case is not interesting.   For dimension $D \ge 5$, no 
function of $G$ is a divergence, so we get  the expected result: scale-invariance and 
almost scale-invariance coincide. Every power of $G$ leads to a scale-invariant Lagrangian, 
and the action   $ I= \int   G^{D/4} \sqrt{-g} d^Dx$ is scale-invariant. \par
So, $D=4$ remains the only interesting case. Here, $G$ represents a divergence, 
and we ask for the set of all almost scale-invariant Lagrangians. The condition 
that the Lagrangian $L=k(G)$ be almost scale-invariant means that there exists 
an $m$ such that
\be 
k( \tilde G) \equiv k( e^{-4c} G ) = e^{mc} k(G) + v(c) \cdot G . 
\ee
 Applying $d/dc$ we
 now get 
$$
-4 k'(e^{-2c} G) \cdot e^{-4c} G = m e^{mc} k(G) + v'(c) \cdot G.
$$
Inserting  $c=0$ and abbreviating $v'(0)$ by $c_2$ we get
\be 
-4 G k'(G)  = m  k(G) + c_2 \cdot G.
\ee
We divide by $G$,   apply $d/dG$ and get 
\be 
-4 \frac{d^2 k}{dG^2} = \frac{d}{dG} \left(  \frac{m k}{G}  \right)
\ee
which is solved by 
\be 
k(G) = c_3  G + c_4  G^{-m/4}
\ee
with  constants $c_3$ and $c_4$. This is just what one expects from the beginning: 
The divergence  $c_3  G$ added to the power-law term $ c_4  G^{-m/4}$.  However, 
eq. (2.15) possesses one further solution besides eq. (2.16): For $m=-4$ one gets 
\be 
k(G) = c_3  G + c_4  G \ln G. 
\ee
The result in eq. (2.17), see eq. (1.3),  was already noted in   \cite{s1}: 
Besides what one would have expected, the action 
 $ I= \int   G \ln G \sqrt{-g} d^4x$  turns out to be  almost scale-invariant.  \par
  An important property,  valid not only for scale-invariant  but also for
almost scale-invariant Lagrangians is the following: If $g_{ij}$ is a vacuum 
solution and $\tilde g_{ij}$ is homothetically related to  $g_{ij}$,  then 
$\tilde g_{ij}$  is also a vacuum solution. For the Lagrangians of type $L=f(R)+k(G)$
 and dimension $D=4$, only  $L= \a R^2$ leads to a scale-invariant action, and 
only $L= \a R^2 + \b G \ln G$  leads to an almost scale-invariant action. This is 
a strong  argument  for a further detailed study of the gravitational  Lagrangian
\be
L_g = \Lambda + R + \a R^2 + \b   G \ln  G +  \gamma C_{ijkl} C^{ijkl}.
\ee
Of course, the term $\gamma C_{ijkl} C^{ijkl}$ -- see  \cite{s34} -- by itself 
has a scale-invariant action, but  we did not consider it in this paper, as it has no 
influence on the field equation within the Friedmann models. The terms $\Lambda$ and 
$R$ are added here since in the weak field limit such terms  appear  effectively. \par
 Einstein's theory of  general relativity has a  scale-invariant Lagrangian, 
but only if the cosmological term is  absent\footnote{or is interpreted as part 
of the matter action}, but not a scale-invariant action. \par
In closing this  section we note that one can construct scale-invariant Lagrangians 
which do not have the form $L=f(R) + k(G)$. One example is 
$$
L = R^{2n} \cdot f(G/R^2)
$$
with  a constant $n$ and an arbitrary (transcendental) function $f$. 

\section{On the Gauss-Bonnet scalar}
\setcounter{equation}{0} 

The field equations for the Lagrangian $L=f(R)$ are given by, see for example 
eq. (2.27) of  \cite{s2}, 
\be 
0 = L_R R_{ij } - g_{ij}L/2 + g_{ij} \Box L_R - \left(L_{R}\right)_{;ij}
  \quad{\rm where} \quad L_R = df/d R \, .
\ee
Since the case $L=f(R)$ has been widely studied we will not go into further details 
here but simply refer the interested reader to the  overview \cite{s2}. \par
    The case when $L=k(G)$ is much less known than the case $L=f(R)$ so we give 
some further details here. For the spatially flat Friedmann metric (given below in 
eq. (4.1)) the Gauss-Bonnet scalar $G$ becomes
\begin{equation} 
G=24 h^2 (h^2 + {\dot h}) = 24 h^4(1+\gamma) ~,
\end{equation}
where $h$ is the Hubble parameter $h = {\dot a}/a$ and $\gamma = {\dot h}/h^2$. 
For  the Lagrangian $k(G)$ with $k_G = dk/dG$, the corresponding
vacuum field equation is given in eq. (3.3) of \cite{s1} as
\bea
0 = \frac{1}{2}g^{ij}k(G) - 2 k_G R R^{ij} + 4 k_G R_k^ i R^{kj}
 - 2 k_G R^{iklm} R^j_{\  klm}  \nonumber  \\
 - 4 k_G R^{iklj} R_{kl} + 2Rk_G^{;ij}
- 2 g^{ij}R \Box k_G - 4 R^{ik} k^{;j}_{G;k}  \nonumber  \\
 - 4 R^{jk} k^{;i}_{G;k} + 4 R^{ij} \Box k_G+ 
4g^{ij} R^{kl} k_{G;kl} - 4 R^{ikjl}  k_{G;kl}  \, .
\eea
See eqs. (A4), (A5) of  \cite{s1}, specialized to the space-time dimension $n=4$: 
\bea
R_{ijkl}=C_{ijkl} + \frac{1}{4} \left( R_{ik} g_{jl} + R_{jl} g_{ik}-
 R_{il} g_{jk}- R_{jk} g_{il} \right ) \nonumber  \\
 - \,  \frac{1}{6}   R \left(   g_{ik} g_{jl} -  g_{il} g_{jk} \right)   \, .
\eea
$ C_{ijkl}$ is the Weyl tensor and  we define $ C^2 = C^{ijkl} C_{ijkl}$ and 
\be
Y^{ij} = R^{iklm} R^j_{\  klm} + 2 R^{iklj} R_{kl}\, .
\ee
With this notation, the first
 (unnumbered) equation
 of the appendix of  \cite{s1} reads
\be
C^{iklm} C_{jklm} = \frac{1}{4} \  \delta^i_j  \ C^2\, .
\ee
Inserting eqs. (3.4) and (3.6) into eq. (3.5) we get
\be
Y^{ij} = \frac{1}{4} g^{ij} C^2 + \frac{1}{6}R^2 g^{ij} - RR^{ij} 
- \frac{1}{2}g^{ij} R_{kl} R^{kl} + 2 R^{ik} R_{\ k}^{j}\, . 
\ee
Further it holds that  
\be
R_{ijkl} R^{ijkl} = C^2 + 2 R_{kl} R^{kl} - \frac{1}{3} R^2
\ee
and
\be
G = C^2 -  2 R_{kl} R^{kl} + \frac{2}{3} R^2 \, .
\ee
With these notations we can rewrite eq. (3.3) as
\bea
0 = \frac{1}{2}g^{ij} \left( k(G) - G k_G \right)  + 2Rk_G^{;ij}
- 2 g^{ij}R \Box k_G - 4 R^{ik} k^{;j}_{G;k}  \nonumber  \\
 - 4 R^{jk} k^{;i}_{G;k} + 4 R^{ij} \Box k_G+ 
4g^{ij} R^{kl} k_{G;kl} - 4 R^{ikjl}  k_{G;kl}  \, .
\eea
Inserting $k(G) = G^n $ into eq. (3.10) we get with $k_G = n G^{n-1}$
\bea
0 = - \frac{n-1}{2}g^{ij} G^n   + 2nR( G^{n-1})^{;ij} - 2n g^{ij}R \Box (G^{n-1})
 - 4n R^{ik} ( G^{n-1})^{;j}_{;k}    \nonumber  \\ - 4n \left( R^{jk} (G^{n-1})^{;i}_{;k}
-  R^{ij} \Box ( G^{n-1}) -
g^{ij} R^{kl} ( G^{n-1})_{;kl} + R^{ikjl} ( G^{n-1})_{;kl} \right)  \, .
\eea
Inserting $k(G) = G \cdot \ln G$ into eq. (3.10) we get with 
$k_G = 1 + \ln G$
\bea
0 = - \frac{1}{2}g^{ij}  G   + 2R(\ln G)^{;ij} - 2 g^{ij}R \Box (\ln G)
 - 4 R^{ik} (\ln G)^{;j}_{;k}    \nonumber  \\ - 4 R^{jk} (\ln G)^{;i}_{;k}
 + 4 R^{ij} \Box (\ln G)+ 4g^{ij} R^{kl} (\ln G)_{;kl} - 4 R^{ikjl} (\ln G)_{;kl}  .
\eea

\section{Cosmological solutions for $\a R^2 + \b G \ln G$}
\setcounter{equation}{0} 

In this section we use the background developed in the previous sections to
give a general study of spatially flat Friedmann space-times
 for almost scale-invariant Lagrangians. In particular we focus one the case
 $L= \alpha R^2 + \beta G \ln G$ which the analysis of section 2 pointed out as an 
important and unique case.  \par
We start by setting up our system and notation.  First, the cosmological metric 
we use is the spatially flat Friedmann space-time given as
\be 
ds^2 = - dt^2 + a^2(t) \left( dx^2+dy^2+dz^2 \right)
\ee
with positive cosmic scale factor $a(t)$. The dot denotes $d/dt$, $h = \dot a/a$  is 
the Hubble parameter, and $R=  6(2h^2 + \dot h)$ is the curvature scalar.\footnote{If 
the metric signature is changed to $(+---)$ then $R$ must be replaced by 
$-R$, whereas $G$ remains the same. In regard to the ambiguities in the sign of 
$R$ often found in the literature we note that we define $R$ such that for the standard 
 2-sphere we always  get $R>0$.}  Without loss of generality we assume
$h\ge 0$. If this is not the case then it is always possible to invert the time 
direction so as to get $h \ge 0$. If $h$ appears in the 
denominator, this automatically  includes the additional assumption, that $h \ne 0$. 
 \footnote{This is not a real restriction, as a constant function $a(t)$ is the trivial 
Minkowski space-time with $h \equiv 0$,  and solutions, where $h(t) =0$ at  isolated 
points $t$ only, can be matched by pieces with $h \ne 0$. In other words:  if $h(t)=0$ at 
isolated points $t$ then these are always connected to regions where $h(t) \ne 0$.} \par
It proves useful to define the function
\be 
\gamma = \dot h / h^2 = - \frac{d}{dt} \left( \frac{1}{h } \right)
 \ee
 which shall be used to replace ${\dot h}$ in subsequent formulas. 
In terms of $\gamma$ we get $R=6 h^2(2+\gamma)$. The deceleration parameter
(i.e. $q=-{\ddot a} a / ({\dot a})^2$) is related to $\gamma$ via $q=-1-\gamma$. \par
Sometimes it proves useful to use $\tau = \ln a$ as an alternative time coordinate. 
With a dash denoting $d/d\tau$, we get with $\dot \tau = h $ the following formula:
\be 
\g' \equiv \frac{d\g}{d\tau}=\frac{d\g}{dt}\cdot \frac{dt}{d\tau}= \frac{\dot \g}{h} 
\ee
We now give some results which will be useful in dealing with the almost 
scale-invariant Lagangians of   the form given in eq. (1.5). First we note that, 
assuming a spatially flat Friedmann metric of the form given in eq. (4.1), that the 
vacuum field equation for a Lagrangian of the form $L = F(G, R)$, where $F$ is a 
function of $R$ and $G$, is (see eq.  (15) of reference \cite{s14})
\be 
0 = G F_G - F - 24 h^3 \dot F_G + 6(\dot h + h^2) F_R - 6h \dot F_R
\ee
where $F_G = \partial F/\partial G$ and $F_R = \partial F/\partial R$.
Eq. (4.4) is the $00$-component of the vacuum field equation for the
Lagrangian $L=F(G,R)$. All other components of the vacuum field equation are 
 fulfilled if eq. (4.4) is valid. For the almost scale-invariant Lagrangian from 
eq. (1.5)  $L = \a R^{2} + \b G \ln G$  we get from  eq. (4.4) 
\be 
0= 3 \a (1+\g) \left(6\g +3\g^2 + 2\g'  \right)-2 \b  \left(1-2\g -3\g^2 - \g' \right) .
\ee
We look for solutions of eq. (4.5) with $\a \b \ne 0$. \par
After some lengthy but straightforward calculations it turned out that for the vacuum 
equation (4.4) following from eq. (1.5) and restricting to the spatially flat Friedmann 
space-time eq. (4.1), no cosmic bounce and no cosmic recollapse is possible; the 
proof was done by inserting a Taylor expansion for  $a(t)$ into the field 
equation (4.4) and to show, that regular local extrema of this function do not exist. 
This result is not very surprising, as one knows this property to be valid already 
for both of the ingredients of eq. (1.5), i.e. for $L=R^2$ and $L=G \ln G$. 
This fact simplifies the calculations as $h(t)$ cannot change its sign, and we do not 
need do match  pieces of different sign of $h$ together.\footnote{For closed Friedmann 
models, however, such a behaviour is possible.}\par 
In a second step, we look for constant values $\gamma \ne 0$ related to the 
scale factor $a(t) = t^p$ representing the self-similar solutions.\footnote{A
space-time is called self-similar, if scale-transformations can  always be 
compensated by  isometries.}
To this end, we insert $\g = -1/p$ into eq. (4.5). Without loss of generality we assume 
$\a = 1/3$ which transforms eq. (4.5)  to 
\be 
0 =    (1 -1/p) \left(-6/p +3/p^2 \right) - 2 \b  \left(1+2/p -3/p^2   \right) \, .
\ee
This equation can be solved for $\b$ by
\be 
\b = -3 (2p-1)/(2p(p+3)).
\ee
As a first estimate we can see the following: 
The leading term in the limit $p \to \infty$  of  eq. (4.6) is $0 = -6/p - 2 \b$. If we insert 
$\b = - 3/p$ into eq. (4.6), then we get qualitatively the following result: A very small 
negative value of the parameter $ \b$ in the Lagrangian 
\be 
L= \frac{1}{3} \cdot R^2 + \b G \ln G
\ee
 leads to the replacement of the exact de Sitter solution from  $L=R^2$, which
 is  a local attractor, to a PLI (power-law inflation)  exact solution  
$a(t) = t^p = t^{-3/\b}$ which also represents a local attractor. This shows how one can
  modify the dynamics from de Sitter to power-law inflation by the addition of 
the $G \ln G$-term. \par
Let us now go more into the details of the system. The limit $\b \to \infty$  in 
eqs. (4.7)/(4.8) is essentially the limit to the Lagrangian $L=G \ln G$ from (4.8) and is
related to the limit $p \to -3$ in eq. (4.7), see the discussion near eq. (1.3) for this case. 
At some points, the Cauchy 
problem fails to be well-posed: For $R=0$, i.e. $\gamma = -2$, the field 
equation following from $L=R^2$ has the property that it is fulfilled by 
 every space-time having  $R \equiv 0$. Thus, the fourth-order 
field equation (3.1)  possessing 10 independent components in the general case, 
now degenerates to one single second order scalar field equation, namely 
$R=0$.\footnote{Sometimes, that value of $R$ where this happens,  is called a 
critical value of  the curvature scalar.} Further, for $G\le  0$, the Lagrangian
$G \ln G$ is not immediately defined; while this might be compensated by
 writing  $G \ln \vert G \vert $ instead for $G <0$, there remains to be a mild
singularity at $G=0$, i.e. for $\gamma = -1$. However, we are in the lucky 
circumstances that for  those  cosmological models where this model may play
 a role, i.e. near de Sitter and near to PLI behaviour, we have $G > 0$ anyhow. 
Therefore we restrict the following discussion to the region $\gamma >-1$. \par
With $3\a =1$, eq. (4.5) can be rewritten as follows: 
\be 
0= 2\g' (1+\b + \g) + (1+\g) \left( 2 \b (3\g - 1) +3\g(2 + \g)  \right)  .
\ee
Here we meet
 the third case
 that the Cauchy problem is not well-posed: namely
 at points where $1+\b + \g = 0$.\footnote{The only exception is the case
$\b =0$, i.e. $L=R^2$, because for this case, eq. (4.9) can be divided by
$1+\g$.} We chose to restrict to that region where $1+\b + \g > 0$. 
For those cases, where $\g' =0$, eq. (4.9) can be solved for $\b$ by 
\be 
\b = - \frac{3 \g (2+ \g)}{2(3\g -1)}.
\ee
Analysing this eq. (4.10) one can see that for a given value $\b$, zero, one or two
related values $\g$ exist. The range of values $\b$, where no $\g$ exists, 
 is the interval 
$$
-2.2 \approx -(\sqrt{7}+4)/3 < \b <  (\sqrt{7}-4)/3 \approx -0.45.   
$$
In the third step we see that eq. (4.9) being of  first order can be analysed 
qualitatively to see the asymptotic behaviour of the solutions: both for the past 
singularity as for the future expansion, the solutions tend to a solution with 
scale factor $a(t) = t^p$. \par
  We count the degrees of freedom as follows:  For a general solution of a fourth-order 
 field equation and one free unknown, in this case $a(t)$,  one expects 4 initial values, 
  namely  $a(0)$ and the first three derivatives of  $a(t)$ at $t=0$. The remaining 
 information is contained in the field equations. However the $00$-component of the
 field equation is a constraint, reducing the order by one. Thus the general solution is 
 expected to have 3 free constants -- one is a $t$-translation, the second is multiplication
  of $t$ by some factor, the third  is multiplication of $a(t)$ by a constant factor. In  this 
 sense, a general solution can be given even if it has no free  parameter.\footnote{It is 
 possible to compensate both the multiplicative   time and the multiplicative space as 
follows: The  multiplication of $t$ by a  non-vanishing factor  can be compensated 
by a scale-transformation of the metric  bringing solutions to solutions  as the field 
equation is scale-invariant, and the  multiplication  of $a$ by a constant factor   
can be compensated by multiplication of   $x$, $y$ and  $z$ by a constant factor. 
This latter possibility is a consequence  of the fact that  the Euclidean 3-space is 
self-similar. Neither  the closed nor the open   Friedmann models  share this 
 property.} \par 
 Let us sum the details: For every fixed $\b >0$, exactly three solutions exist. 
 They can be described as follows: Find that value $\g (\b)$ which has 
$0 < \g (\b) < 1/3$ and solves eq. (4.10). Then the solution with $\g = \g (\b)$
 is just the self-similar solution $a(t) = t^p$ with $p = - 1/\g < -3$ discussed 
already earlier, see eq. (4.7).\footnote{Of course, for $p<0$, writing simply 
$a(t) = t^p$ may be misleading, because then  $h <0$ appears; but we believe
 that  it is sufficiently clear that  for $p<0$, writing simply 
$a(t) = t^p$ means $a(t) = (t_0 - t)^p$ with $t < t_0$. }
 The second solution has $\g > \g (\b)$ and $\g' <0$ throughout, starts from an 
initial singularity  with $\g \to \infty$ and attracts the self-similar solution 
$a(t) = t^p$ which, of course, reaches $a \to \infty$ in a finite time $t$. 
 The third solution has $-1 < \g < \g (\b)$ and $\g' >0$ throughout, starts from an 
initial singularity  of type  $a(t) = t $  and also  attracts the self-similar solution 
$a(t) = t^p$ for $a \to \infty$. \par
For $ -0.45 \approx  (\sqrt{7}-4)/3 \le  \b < 0$ we have essentially the same
behaviour as in the previous case: One self-similar solution $a(t) = t^p$, but
now with $p>0$, and two other ones, both having  $a(t) = t^p$ as attractor
 for $t \to \infty$. For the remaining negative values of $\b$ the instabilities 
become  more serious, and the behaviour of the solution has several 
different types of singularities. We conclude that probably the physically sensible 
range of  our model is in the interval $ \b \ge  (\sqrt{7}-4)/3 $. \par
In closing this section we show that it is possible to completely integrate eq. (4.5) 
for the special case when $\alpha=0$; i.e. when $L=\beta G \ln G$ one can integrate 
the system completely to the point where one has an explicit form for the scale 
factor $a(t)$. This special case can be thought of as the limit where $\beta >> \alpha$
 so that system is dominated by the Gauss-Bonnet logarithmic term. In this case 
taking $\alpha =0$ eq. (4.5) becomes $\g'  = A \gamma ^2 +B \gamma +C$ 
where the prime indicates differentiation with respect to $\tau$ and where $A, B, C$ are
\begin{equation}
\label{4.11}
A=-3 ~~;~~~~ B= -2 ~~;~~~~ C=1 ~.
\end{equation}
However we will carry through the calculation until almost the end using general 
$A, B, C$ since in this way the analysis can also be applied to the special
 Lagrangians $L=R^{2n}$ and $L=G^n$ which also lead to an equation for $\gamma$ 
of the form  $  \g' = A \gamma ^2 +B \gamma +C$. Integrating this 
equation for $\gamma$ yields
\begin{equation}
\label{4.12}
\tau = \int \frac{d \gamma}{A \gamma^2 + B \gamma + C} = 
\frac{2}{D} {\rm arctan}  \left( \frac{B+2 A \gamma}{D}\right)
\end{equation}
where $D=\sqrt{-B^2 +4 A C}$. Inverting eq. (\ref{4.11}) gives
\begin{equation}
\label{4.13}
\gamma (\tau ) = \frac{D}{2A} \tan \left( \frac{D \tau}{2} \right) - \frac{B}{2A}~,
\end{equation}
Now taking into account the form of $A, B, C$ from eq. (\ref{4.11})  we find that 
$D =\sqrt{-B^2 +4 A C}$ is imaginary. Thus in eq. (\ref{4.13}) we replace $D$ with
 $i D_1$ where $D_1 =\sqrt{B^2- 4 A  C}$ and taking into account that 
$\tan(i D_1) = i {\rm  tanh} (D_1 )$ we find that  eq. (\ref{4.13}) becomes
\begin{equation}
\label{4.14}
\gamma (\tau ) =
 -\frac{D_1}{2A} {\rm \tanh} \left( \frac{D_1 \tau}{2} \right) - \frac{B}{2A}~.
\end{equation}
Next, we use this $\gamma(\tau )$ from eq. (\ref{4.14}) to  solve the equation for 
$h={\dot a}/a$
 (where the overdot
 is differentiation 
with respect to $t$). 
First we
 note that 
\begin{equation}
\label{4.15}
\gamma (\tau) = -\frac{d}{dt} \left( \frac{1}{h} \right) = 
-\frac{d \tau}{dt} \frac{d}{d \tau} \left( \frac{1}{h} \right) =
\frac{1}{h} \frac{dh}{d \tau} ~.
\end{equation}
We have used $h= d \tau / dt$ in arriving at the final result. Now we integrate 
eq. (\ref{4.15}) for $\gamma (\tau)$ from eq. (\ref{4.14}) which yields
\begin{equation}
\label{4.16}
\ln (h(\tau) ) = -\frac{1}{A} \ln \left[ \cosh \left( \frac{D_1 \tau}{2} 
\right) \right] - \frac{B}{2A} \tau ~,
\end{equation}
or solving for $h ( \tau )$ gives
\begin{equation}
\label{4.17}
h(\tau)  = \left[ \cosh \left( \frac{D_1 \tau}{2} \right) \right]^{-1/A} 
\exp \left( - \frac{B}{2A} \tau \right) ~.
\end{equation}
Now using $h (t) = {\dot a(t)}/a(t)$ for the left hand side and $\tau = \ln [a(t)]$ in
 the right hand side of  eq. (\ref{4.17}) gives
\begin{eqnarray}
\label{4.18}
\frac{{\dot a}}{a}  &=& \left[ \frac{1}{2} \left(  \exp \left( \frac{D_1 
\ln (a)}{2} \right) 
+ \exp \left( - \frac{D_1 \ln(a)}{2} \right) \right) \right]^{-1/A} \exp \left( 
- \frac{B}{2A} \ln (a) \right) \nonumber \\
&=& \left( \frac{1}{2} \right) ^{-1/A} \left( a^{(D_1+B)/2} + 
a^{(-D_1+B)/2} \right) ^{-1/A}   \nonumber \\ 
&=& \left( \frac{1}{2} \right) ^{1/3} \left( a + a^{-3} \right) ^{1/3}  ~, 
\end{eqnarray}
where in the last line we have inserted the specific values of $A, B, C, D_1$ for this
 cases when $L=G \ln G$ i.e. $A=-3$, $B=-2$, $C=1$ and $D_1=4$.
 Finally integrating eq. (\ref{4.18}) gives
\begin{equation}
\label{4.19}
\int dt = t  = 2^{1/3}  \int   a^{-1} \left( a + a^{-3} \right) ^{-1/3} da  
= 2^{1/3} a(t) ~ {_2 F _1} \left( \frac{1}{4} , \frac{1}{3} ; \frac{5}{4} ;  
-a ^4 (t) \right)  +k
\end{equation}
where $k$ is an integration constant and $_2 F _1 (a,b;c;z)$ is the hypergeometric 
function. One should now solve eq. (\ref{4.19}) for $a(t)$ to obtain the scale factor 
as a function of $t$. Or a simpler method -- given the presence of $_2 F _1$ in eq. 
(\ref{4.19}) -- is to plot $t$ versus $a$ and then flip the graph about the line $t=a$ 
thus graphically giving $a(t)$.  If one does this then one finds that $a(t)$ is an 
exponentially growing function of $t$, thus having a term like $G \ln G$ might 
be a way to obtain an early inflationary stage to the Universe.  

\section{Discussion}
\setcounter{equation}{0} 

In section 2   we had introduced the notions of scale-invariant and almost 
scale-invariant Lagrangians. For the set of Lagrangians $L=f(R) + k(G)$
 we found out that $L$ is scale-invariant if
 \be
L = \a R^{2n} + \b G^n
\ee
with constants $\a$, $\b$, and $n$.   This is a closed set of functions. If we add
the divergence $\gamma G$ with constant $\gamma$, we arrive trivially at the
following set of almost scale-invariant Lagrangians
 \be
L = \a R^{2n} + \b G^n+ \gamma G.
\ee
Of course, both Lagrangians give rise to the same set of field equations, but the
4-parameter set eq. (5.2) fails to be a closed set of functions. To see this, 
we take the limit $n \to 1$ in eq. (5.2) while $\a$ remains constant 
and $ \b = -  \g = 1/(n-1)$. We get
$$
\lim_{n \to 1} \  \a R^{2n} +  G \cdot \frac{G^{n-1}-1}{n-1} =\a R^2 + G \ln G.
$$
In arriving at this result we have defined $n-1=\e$ and $G=e^z$ and used
$$
\lim_{n \to 1} \  \frac{G^{n-1} - 1}{n-1}  = \lim_{\e \to 0} \  \frac{G^\e -1}{\e}
= \lim_{\e \to 0} \ \frac{e^{\e z} -1}{\e} = z = \ln G. 
$$
Result: Besides the trivial almost scale-invariant Lagrangians, as defined by eq. (5.2),
 the class of almost scale-invariant Lagrangians  include also the 
 Lagrangians of the  form\footnote{We note that, since the argument of the logarithm 
in eq. (5.3) should  be positive and dimensionless we should replace $\ln G$ 
by $\ln (G/G_0 )$   where $G_0$ is some  constant $G_0 \ne 0$. }  
\be 
L = \a R^{2} + \b G \ln G + \g G .
\ee
 Of course the inclusion of matter will change the behaviour of the 
cosmological solutions discussed in this paper, but in the early stages of the 
Universe, with matter in the form of dust or radiation, the behaviour of the
solutions above will only be marginally modified by the presence of the matter. 
The behaviour described in this paper will thus essentially correctly describe the
 dynamics.  \par 
Finally, let us reformulate one of the key results of this work given in section 
4 which is closely related to analogous
calculations done for higher dimensions in \cite{s998}: For $p>0$,  the Lagrangian 
$$
L= \frac{1}{3} \cdot R^2 - \frac{3}{p+3} \cdot \frac{2p-1}{2p}  \cdot G \ln G
$$
 has the spatially flat Friedmann with scale factor  $a(t) = t^p$ as exact vacuum solution.
For large  values $p$ this is a local attractor solution and it represents a model for
power-law inflation.

\section*{Acknowledgements}%

Useful comments by S. Deser, Q. Exirifard, A. de la Cruz-Dombriz  and S. Odintsov 
are gratefully acknowledged. DS acknowledges the support of  a DAAD  (Deutscher
Akademischer Austauschdienst) grant to do research at  Universit\"at  Potsdam.

\end{document}